# Federated Learning Approach for Lifetime Prediction of Semiconductor Lasers

**Khouloud Abdelli[1,2], Helmut Grießer[1], and Stephan Pachnicke[2]**
[1]*ADVA Optical Networking SE, Fraunhoferstr. 9a, 82152 Munich/Martinsried, Germany*
[2] *Christian-Albrechts-Universität zu Kiel, Kaiserstr. 2, 24143 Kiel, Germany*
E-mail: KAbdelli@adva.com

**Abstract:** A new privacy-preserving federated learning framework allowing laser manufacturers to collaboratively build a robust ML-based laser lifetime prediction model, is proposed. It achieves a mean absolute error of 0.1 years and a significant performance improvement. © 2022 The Author(s)

## 1. Introduction

Despite nearly 60 years of field service, the reliability of semiconductor laser is still a challenging issue for the optical fiber communications field requiring both high performance and long-term reliability (i.e., a mean time to failure higher than $10^5$ hours) [1]. The qualification of semiconductor laser reliability is performed using laboratory data, derived from accelerated aging tests conducted under stressed conditions such as high temperature and/or device drive current. This speeds up the degradation and thereby shortens the time to failure of the device, otherwise it takes many years to collect meaningful field lifetime data. However, estimating the lifetime of semiconductor lasers by extrapolating aging results [2] is not very accurate [1]. Recently, machine learning (ML) based approaches have emerged as powerful tools for solving prognostic problems for semiconductor lasers [3]. Still, aging tests are carried out only for few devices due to the high effort and costs, limiting the amount of training data and thus prediction accuracy and reliability of the ML models. Federated learning (FL) is a promising candidate to tackle the aforementioned issue by enabling the development of a global ML model using datasets owned by many parties without revealing their business-confidential data. In this respect, we presented a secure FL framework to predict the maintenance work for semiconductor lasers running in an optical network [4]. After the global ML model is deployed, each laser manufacturer benefits by receiving the personalized maintenance report on their hardware failure rate.

In this paper, we propose a FL approach, which allows many semiconductor laser manufacturers (i.e. clients) to collaborate in order to develop a robust common ML model for semiconductor laser lifetime prediction before their deployments in optical networks under the coordination of a central server without sharing their local private data with others. After the end of the training phase, the built global ML model is sent to all the clients, which use it for enhancing the reliability qualification of their products, and to save the costs of performing time consuming aging tests while ensuring high prediction accuracy. The ML model is an attention-based deep learning approach leveraging the benefits of gated recurrent unit (GRU) networks in processing sequential data, the capability of attention mechanisms to learn the most important features and the inclusion of statistical features in order to improve the prediction accuracy. To verify the effectiveness of the proposed FL framework, we use vertical-cavity surface-emitting laser (VCSEL) reliability data derived from accelerated aging tests conducted under different operating conditions. The experimental results show that our framework achieves a good lifetime prediction capability while preserving privacy, and provides an improvement in performance and robustness when compared with models trained at a single client's premises using that client's data.

## 2. Experiments

*2.1 Proposed federated learning framework* We consider a semiconductor laser lifetime prediction framework that assumes $N$ laser manufactures (i.e. clients) for collaborative training of a global ML model under the control of an aggregator server ($S$), while keeping every client's data private as shown in Fig.1. $S$ starts by randomly initializing the global model and notifies each client ($C_i$) to train a globally shared model using his local data $D_i$. Afterwards, $S$ distributes the global model parameter vector $w_t$ to the participating clients. Then, each $C_i$ trains locally the ML model with his data consisting of $n_i$ samples, independently. The local model parameters are updated by minimizing the loss function over $n_i$, starting from the global model parameter vector $w_t$ shared by $S$ and using an Adam optimizer. At the end of the local training phase, the locally trained parameters ($w_{t+1}^i$) are sent to $S$ and aggregated by computing a weighted average resulting in the global model parameter for the next round ($w_{t+1}$) [5]:

$$w_{t+1} = \sum_1^N \frac{n_i}{\sum_i n_i} w_{t+1}^i \quad (1)$$

The update of the local and global models is repeated for a certain number of iterations (communication rounds $N_{round}$) until the global loss function converges.

*2.2 Machine learning model architecture*

Fig. 2 illustrates the structure of the proposed local model for the semiconductor laser lifetime prediction. The proposed approach is an attention-based deep learning model that makes full use of the fusion of the sequential features learned by the attention-based GRU layers, the statistical features characterizing the degradation trend, namely kurtosis ($\beta$) and skewness ($\delta$), and the operating conditions temperature $T$ and laser current $I$, impacting the degradation, in order to improve the prediction accuracy. The sequential input [$P_0, P_1, \ldots P_8$] is fed to 2 GRU layers composed of 64 and 32 cells, respectively, to learn the relevant sequential features [$h_0, h_1, \ldots h_8$]. Then, the attention layer assigns to each extracted feature $h_i$ a weight (i.e. attention score) $\alpha_i$, which is calculated as follows:

$$\alpha_i = \text{softmax}(w^T \tanh(W_h h_i)) \quad (2)$$

where $W_h$ and $w$ denote weight matrices. Softmax is used to normalize $\alpha_i$ and to ensure that $\alpha_i \geq 0$, and $\sum_i \alpha_i = 1$. The different computed weights $\alpha_i$ are then aggregated to obtain a weighted feature vector (i.e. attention context vector) $c$, computed as $\sum_i \alpha_i h_i$, which captures the relevant information to improve the performance of the model [6]. In parallel, the statistical features combined with $T$ and $I$ are given to a fully connected layer containing 64 neurons. Afterwards, the fully connected layer merges the learned feature outputs, and the most important sequential features are extracted by the attention layer. The fused features are then fed to two fully connected layers with 64 and 32 neurons, respectively, followed by a dropout layer to avoid overfitting, that outputs the time to failure of the device ($TTF$). The whole network is simultaneously trained by minimizing the cost function (mean square error (MSE)) adjusted by adopting an Adam optimizer. Note that the global model has the same architecture as the local model, and the loss function of the global model is computed as the weighted average of the local models' losses.

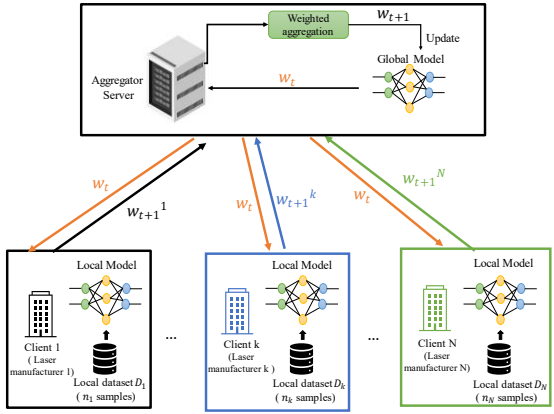 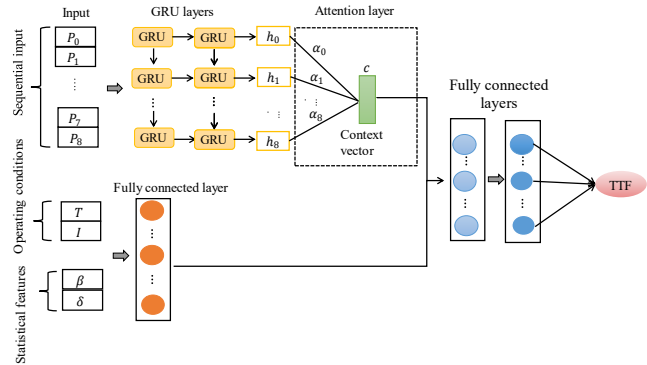

Fig. 1. Federated laser lifetime prediction learning architecture. Fig. 2. Architecture of the proposed ML local model.

*2.3 Experimental data*

The proposed framework is validated using experimental data derived from several accelerated aging tests performed for VCSELs with various oxide aperture sizes and with varying controlled operating conditions such as temperature $T$ and laser current $I$. The aging tests are carried out under high temperature (50°C $\leq T \leq$ 150°C) to strongly increase the laser degradation and thereby speed up the device failure. The output power (i.e. degradation parameter) is monitored under constant current operation. The duration of the aging tests is either 3,500 h or 15,000 h. $TTF$ is defined as the time at which the output power has decreased by 1 dB (20%) of its initial value. In total, a dataset of 3,397 samples incorporating the sequences of monitored output power measurements combined with $\beta$, $\delta$, $T$ and $I$, is built. $TTF$ is assigned to each sample. The said data is then normalized and divided into a training data (comprising of 90% of the samples used for training the local models) and a test dataset (the remaining 10% for testing). The training data is then split into $N = 8$ clients with different parts. Note that the data owned by each client is derived from accelerated aging tests of VCSELs conducted under operating conditions that are different from the other clients' data, leading to an heterogeneous federated setting.

## 3. Results and Discussions

The performance of the proposed FL framework is compared to two baseline models including a model trained by applying a traditional centralized approach and a locally trained model without participating in the FL approach (i.e. localized model). The centralized approach is trained with the data from all the clients, which is collected and stored at a single server. The localized model is trained on the client's premises without model sharing during the training procedure. The different approaches are evaluated using as evaluation metrics the root mean square error (RMSE), the standard deviation (SDEV) and the mean absolute error (MAE). Figure 3 (a) illustrates the statistics of the performance achieved by the localized models trained with the local data of each client. The average scores of RMSE, SDEV and MAE are 0.66 years, 0.48 years, and 0.46 years, respectively. The best performing localized model is the model trained by the larger local data owned by the client $C_8$ (541 samples), whereas the worst performance is yielded by the model trained with the smaller local data (226 samples). For the localized approach, the best model achieved by $C_8$ is selected as a reference to be compared with the other approaches. Note that the improvement in performance

($\Delta$) achieved by the FL over the localized model is calculated as follows for an evaluation metric $m$:

$$\Delta\ (\%) = \left(1 - \frac{m_{FL}}{m_{localized}}\right) \times 100 \quad \forall\ m\ \in \{RMSE, SDEV, MAE\} \quad (3)$$

The results of the comparison shown in Fig. 3 (b) demonstrate that first the proposed FL model outperforms the localized approach by providing 64%, 57.5% and 74.35% improvements in RMSE, SDEV and MAE metrics, respectively, and second that the FL approach achieves similar performance as the centralized approach while ensuring data privacy. Figure 3 (c) shows that from 70 communication rounds the FL approach yields the same MAE, achieved by the centralized model tested with unseen data, and that the FL model framework has good convergence and stability.

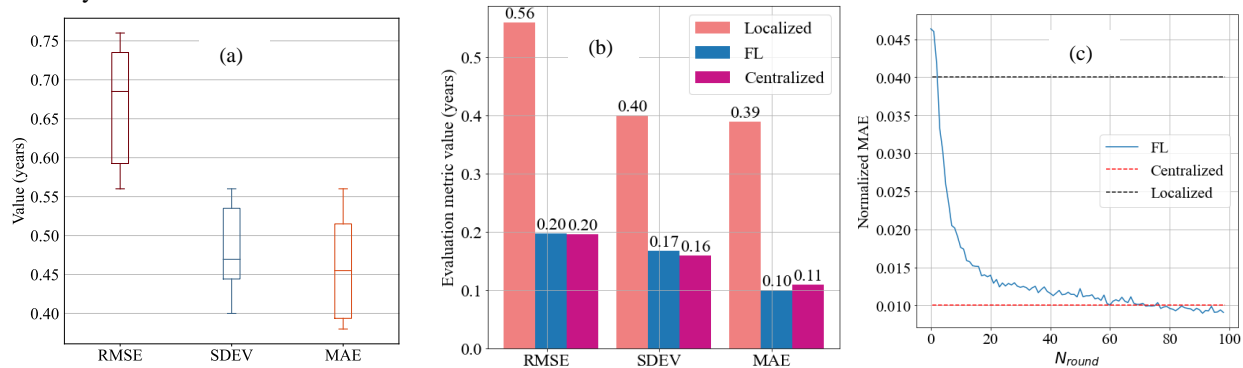

Fig. 3. Performance evaluation. (a) Performance statistics of the localized models, (b) Comparison of the federated (FL), centralized and localized approaches, (c) Normalized mean absolute error of the FL approach with number of communication rounds.

Fig. 4 illustrates that the predicted TTFs by the FL approach are close to the true TTF values, which proves the effectiveness of the FL framework in accurately estimating the laser lifetime. The results presented in Fig. 5 show that with an increasing number of clients the performance of the FL model decreases. Splitting the same dataset between more clients reduces the amount of local data, slowing down the convergence of the global model and impacting the performance.

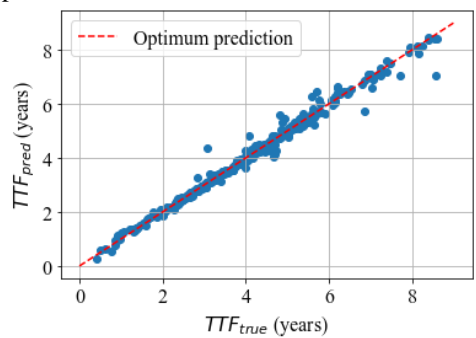
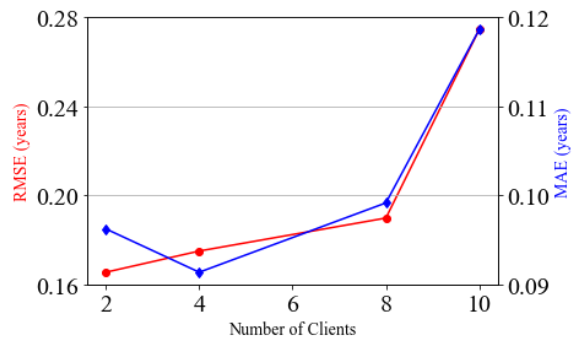

Fig. 4. Predicted TTFs by FL approach vs. actual TTFs.

Fig. 5. Impact of the number of clients on FL model performance.

## 4. Conclusion

We proposed a federated learning framework for semiconductor laser lifetime prediction. The proposed approach has been evaluated using VCSEL lifetime data. The results demonstrate that the presented FL approach achieves a good prediction capability (a MAE of 0.1 years) similar to the one yielded by the centralized model, with privacy well-preserved, and provides a significant performance improvement across all participating clients.

**Acknowledgements**
This work has been performed in the framework of the CELTIC-NEXT project AI-NET PROTECT (Project ID C2019/3-4), and it is partly funded by the German Federal Ministry of Education and Research (FKZ16KIS1279K).